\documentclass{scrartcl}
\usepackage[utf8]{inputenc}
\usepackage{setspace}
\usepackage{float}
\usepackage{newfloat}
\usepackage[hidelinks, breaklinks=true]{hyperref}
\usepackage[hyphenbreaks]{breakurl}
\usepackage{longtable}
\usepackage[labelsep=period, singlelinecheck=false, justification=RaggedRight, format=plain]{caption}
\usepackage[flushleft]{threeparttablex}
\usepackage{pdflscape}
\usepackage{appendix}
\usepackage[affil-it]{authblk}
\usepackage[natbibapa]{apacite}
\usepackage{cite}

\setkomafont{disposition}{\normalcolor\bfseries}

\newcommand\blfootnote[1]{%
  \begingroup
  \renewcommand\thefootnote{}\footnote{#1}%
  \addtocounter{footnote}{-1}%
  \endgroup
}

\title{The Second-Level Smartphone Divide: A Typology of Smartphone Usage Based on Frequency of Use, Skills, and Types of Activities}

\author[1]{Alexander Wenz}
\author[1]{Florian Keusch}

\affil[1]{\footnotesize School of Social Sciences, University of Mannheim, Germany}

\date{}

\begin{document}
\maketitle

\begin{abstract}
\noindent This paper examines inequalities in the usage of smartphone technology based on five samples of smartphone owners collected in Germany and Austria between 2016 and 2020. We identify six distinct types of smartphone users by conducting latent class analyses that classify individuals based on their frequency of smartphone use, self-rated smartphone skills, and activities carried out on their smartphone. The results show that the smartphone usage types differ significantly by sociodemographic and smartphone-related characteristics: The types reflecting more frequent and diverse smartphone use are younger, have higher levels of educational attainment, and are more likely to use an iPhone. Overall, the composition of the latent classes and their characteristics are robust across samples and time. \blfootnote{\textbf{Acknowledgements:} This work was supported by the German Research Foundation (DFG) through the Collaborative Research Center SFB 884 “Political Economy of Reforms” (Project A8) [139943784 to Annelies Blom, Florian Keusch, and Frauke Kreuter].}
\end{abstract}

\onehalfspacing
\section{Introduction}
Smartphones have become an integral part of our daily lives, with 85\% of U.S. adults owning a smartphone in 2021, compared to only 35\% of adults a decade ago \citep{PewResearchCenter2021}. A similar increase in smartphone ownership has been reported across other countries. For example, \citet{Eurostat2021} shows that the proportion of people in Germany aged 16 and older who use a smartphone or other mobile phone to access the Internet increased from 15\% in 2011 to 75\% in 2019. While this proportion is still lower in emerging economies, smartphone ownership rates are also rapidly increasing in these countries \citep{Silver2019}.

Despite the increasing smartphone ownership in the general population, inequalities in smartphone access continue to exist \citep{VanDeursen2019}. Recent data from early 2021 show that 96\% of U.S. adults under the age of 30 but only 61\% of adults aged 65 and older own a smartphone. Smartphone ownership increases with household income and educational attainment, and U.S. adults living in rural areas are less likely to own a smartphone than adults living in urban or suburban areas \citep{PewResearchCenter2021}, with similar patterns in other Western countries \citep{Eurostat2021, Keusch2020}.

Beyond disparities in digital access, digital divide scholars have increasingly paid attention to inequalities related to digital skills and the usage of digital technologies, which is often referred to as the second-level digital divide \citep{DiMaggio2004, Hargittai2002}. Among people with Internet access, considerable variation has been found in how effectively they can use and take advantage of the Internet, exacerbating existing digital inequalities in terms of access \citep{VanDijk2020}. While a growing body of research has addressed these usage inequalities with a focus on general Internet use, surprisingly little is known in the digital divide literature about inequalities in the general population related to smartphone usage. Studying the second-level smartphone divide becomes increasingly relevant given that people with low levels of smartphone skills and usage are likely to be excluded from an increasing array of outcomes. Smartphones, for example, are required to use popular communication apps, such as WhatsApp \citep{Taipale2018}, have increasingly been used for medical apps to prevent and treat chronic diseases \citep{Gustafson2014}, and have been used for contact tracing in the COVID-19 pandemic \citep{Blom2021}. In addition, smartphones become increasingly relevant in the context of commercial activities, for example to use online banking and make purchases via apps \citep{Gross2015, Singh2020}, or make mobile payments \citep{Gerpott2017}. Although smartphone technologies have been fast changing, with an increasing number of functionalities and sensors, there has only been limited research about whether inequalities in smartphone usage have changed over time. This paper fills this research gap by addressing the following research questions:

\begin{description}
    \itemsep0em 
    \item \textbf{RQ1.} Which smartphone usage types can be identified among smartphone owners?
    \item \textbf{RQ2.} How do the smartphone usage types differ in sociodemographic and smartphone-related characteristics?
    \item \textbf{RQ3.} How stable are the smartphone usage types across time and samples?
\end{description}

The paper is organized as follows. In the next section, we review the related literature on the second-level digital divide, with a focus on existing typologies of Internet and digital media use as well as correlates of use. To answer our research questions, we present survey data from five samples of smartphone owners collected in Germany and Austria between 2016 and 2020. We identify smartphone usage types by conducting a latent class analysis that classifies individuals based on their frequency of smartphone use, self-rated smartphone skills, and activities carried out on their smartphone. We then examine how these usage types differ in sociodemographic and smartphone-related characteristics. Finally, we analyze whether the smartphone usage types have changed across the samples collected between 2016 and 2020. We conclude the paper with a discussion of our findings and the implications for digital divide research.

\section{Background}
\subsection{Typologies of Internet and digital media use}
Over the last two decades, an extensive number of typologies have been constructed to conceptualize and classify the large variation in Internet and digital media use \citep{Blank2014}. In one of the earlier studies, \citet{Selwyn2005}, for example, interviewed 1,001 adults in four regions in the United Kingdom and identified four categories of Internet users based on their frequency and type of Internet use: 1) broad frequent users (13\% of the sample), 2) narrow frequent users (18\%), 3) occasional users (11\%), and 4) non-users (58\%). In a more recent study, \citet{Reisdorf2017} use data from the Oxford Internet Survey, a nationally representative face-to-face survey of individuals aged 14 and older in Great Britain, to identify five types of Internet users: 1) broad users (27\% of the population), 2) regular users (30\%), 3) low users (21\%), 4) non-users (18\%), and 4) ex-users (4\%).

To create the usage typologies, prior studies often classified Internet users based on their amount, variety, and types of use \citep{Blank2014, Holmes2011, Horrigan2007, Livingstone2007, Reisdorf2017, Selwyn2005, VanDeursen2014, Zillien2009}. Whereas amount of use is measured as frequency of going online (e.g., number of hours per day), frequency of engaging in different online activities, or total years of Internet use, variety of use is measured as the number of activities carried out online. Types of use, in turn, are measured by nominal categories that represent different online activities, such as information seeking and communication.

While a large number of typologies have been proposed for general Internet usage, smartphone usage has received less attention in prior research on the digital divide. In a few of the existing studies, mobile phone and smartphone use has been employed as an indicator for general Internet usage \citep{Herzing2019, Horrigan2007, Yates2020} but has not been studied on its own. Two related areas of research where smartphone user typologies have previously been developed, however, are smartphone addiction research \citep{Bian2015, Elhai2018, Kim2016} and marketing research \citep{Calvo-Porral2020, Chen2019, DeCanio2016, Hamka2014, Petrovcic2018, Sell2014}.

In the area of smartphone addiction research, \citet{Elhai2018}, for example, examined the relationship between usage patterns and problematic smartphone use in a survey of 296 college students in the U.S. To classify the students into types of smartphone users, they conducted a latent class analysis based on the students’ self-reported frequency of engaging in eleven different activities on their smartphone, such as voice/video calls, texting/instant messaging, and using social media. The authors identified two classes of smartphone users, including heavy users who use their phone for a large variety of activities, and light users who particularly make use of social media, audio entertainment as well as photo- and video-taking. 

In the marketing literature, typologies of smartphone users have been proposed in the context of customer segmentation research. For example, \citet{DeCanio2016} collected survey data from 264 smartphone users in Italy to conduct a cluster analysis based on their self-reported use of ten different smartphone functions. The authors identified five types of smartphone users which they called “unfriendly users” (who do not use any of the ten smartphone functions), “utility users” (who mainly use their smartphone for voice calling and information seeking), “gamers” (who mainly use their smartphone for games), “moderator users” (who use all of the ten smartphone functions, but information seeking in particular), and “supersmartphoners” (who mainly use their smartphone for taking photos and videos, social media, and staying in contact with others, such as video calling).

Using a different measurement approach, \citet{Hamka2014} conducted a latent class analysis based on log files from 129 smartphone users who installed a tracking app on their device for at least two weeks. The authors identified six types of smartphone users, including “application ignorant users” (who visit a small number of URLs in the mobile browser and use a small number of apps per day), “basic application users” (who visit a small number of URLs but use a medium number of apps), “average application users” (who visit a medium number of URLs and use a medium number of apps), “information seekers” (who visit a large number of URLs but use a small number of apps), “app savvy users” (who visit a large number of URLs, use an extensive number of apps, and install a large number of new apps), and “high utility users” (who visit an extensive number of URLs, use an extensive number of apps, but install a small number of new apps).

\subsection{Correlates of Internet and digital media use}
Compared to the large body of research on correlates of general Internet usage \citep{Blank2014, VanDeursen2014}, prior research on smartphone usage has been relatively sparse. Frequency and variety of smartphone use have been found to decrease with age \citep{Andone2016, Fortunati2014, Serrano-Cinca2018}. With regard to the types of activities carried out on the smartphone, younger individuals are more likely to use their smartphone for entertainment and games, social interaction, and commercial activities \citep{Andone2016, Kongaut2016}. However, no significant age effect has been found for reading and writing emails on the smartphone \citep{Kongaut2016}.

With regard to gender, previous research has shown that women use smartphones for a larger amount of time and visit a larger number of websites when using a browser on their smartphone \citep{Andone2016, Cotten2009, Roberts2014, Wang2018}. In contrast, men use a larger number of apps on their smartphone and generate a larger volume of uploaded and downloaded data traffic via the cellular network \citep{Wang2018, Zhao2020}. The evidence for gender differences in the variety of smartphone use, however, is mixed \citep{Cotten2009, Fortunati2014, Mascheroni2016, Serrano-Cinca2018, Wang2018, Zhao2020}. Women are more likely to use their smartphone for social interaction and taking photos whereas men, in turn, are more likely to read or watch news on their smartphone \citep{Andone2016, Cotten2009, Kim2016, Kongaut2016, Mascheroni2016, Roberts2014, Zhao2020}. Mixed findings were reported for other smartphone activities, including entertainment and games, commercial activities, and reading and writing emails \citep{Andone2016, Cotten2009, Kongaut2016, Mascheroni2016, Roberts2014, Wang2018, Zhao2020}.

Individuals with higher levels of educational attainment were shown to use their smartphone for a larger number of activities \citep{Fortunati2014, Serrano-Cinca2018}. They are more likely to use their smartphone for entertainment and games, social interaction, commercial activities, and reading and writing emails \citep{Kongaut2016}.

\subsection{Internet and digital media use over time}
Prior research about how inequalities in Internet and digital media use have changed over time is even further limited. In one of the few studies, \citet{VanDeursen2015} used repeated cross-sectional data from the general population in the Netherlands to examine how types of Internet activities and their sociodemographic correlates have changed between 2010 and 2013. They found that individuals increasingly used the Internet for news consumption, commercial activities, and social interaction whereas information seeking, the most popular Internet activity, remained on a constant level over this period. With regard to sociodemographic correlates, they showed that age differences decreased for news consumption, commercial activities, and social interaction on the Internet. In contrast, differences by educational attainment increased for news consumption and information seeking: Individuals with high educational attainment, compared to those with medium educational attainment, were in 2013 even more likely to read or watch news or search for information on the Internet than in 2010. Finally, the differences by gender in types of Internet activities were mostly consistent between 2010 and 2013.

To investigate how app usage behavior has changed over time, a few studies have also relied on log files collected from smartphone users who installed a tracking app on their device.  For example, \citet{Kim2019} collected data from 139,935 iPhone users around the world between 2012 and 2016 and found that the use of social media apps and entertainment apps, including photo, video, and gaming apps, increased over this period while the use of productivity apps decreased. Similarly, \citet{Li2020} used data from 1,465 Android smartphone users around the world, collected between 2012 and 2017, to study longitudinal changes in app usage. They found that the diversity of app usage increased over the five-year period while the number of apps used increased between 2012 and 2014 and decreased between 2014 and 2017.

In this study, we aim to extend the prior literature on the second-level digital divide by developing a typology of smartphone usage that is based on frequency of use, self-rated skills, and types of activities carried out on the smartphone. Compared to previous smartphone usage typologies, we use multiple samples that are representative of the general population and have a larger sample size, allowing us to take a more nuanced perspective on the different usage patterns and to assess how stable the typology is across samples and time. Finally, our study aims to add to the relatively small body of research on sociodemographic correlates of smartphone usage.

\section{Data and Methods}
\subsection{Samples}
To answer our research questions, we collected survey data from five samples of smartphone owners in Germany and Austria between 2016 and 2020. Measuring the same constructs in different samples allows us to examine change in smartphone usage types over time, but also to replicate our findings under different conditions.

\textbf{Sample 1.} The first sample comes from a web survey conducted among members of a German nonprobability online panel in December 2016 \citep{Keusch2019b}. A total of 9,000 panel members received an email invitation and 3,144 individuals started the survey. Quotas for gender and age were used to get a sample of sufficient diversity, and only smartphone owners were eligible for the study. Of the panel members who started the survey, 404 were screened out because of the quotas, and 32 were screened out because they did not own a smartphone. Of the 2,708 remaining respondents, 61 dropped out of the survey (2.2\%) while 24 had duplicated IDs and were removed from the dataset, resulting in an analysis sample of N=2,623.

\textbf{Sample 2.} The second sample comes from Wave 32 of the German Internet Panel (GIP) conducted in November 2017 \citep{Blom2018}. The GIP is a probability-based online panel of the German general population aged 16-75 \citep{Blom2015}. Sample members were recruited face-to-face in 2012 and 2014. Individuals without computer and/or Internet access were provided with computing equipment and broadband Internet \citep{Blom2017}. Every two months panel members are invited via email to complete web surveys on political and economic attitudes. A total of 2,648 panel members answered the survey, resulting in a completion rate of 53.3\% and a cumulative response rate of 10.9\%. Only smartphone owners were asked the relevant questions for our analysis, resulting in an analysis sample of N=2,186.

\textbf{Sample 3.} The third sample comes from a web survey conducted among members of another German nonprobability online panel in December 2017 \citep{Keusch2019a}. Panel members were invited through a survey-router system to participate in the survey, and 1,398 individuals started the survey. Quotas for gender and age were again used, and only smartphone owners were eligible for the study. Of the panel members who started the survey, 82 were screened out because of the quotas or because they did not own a smartphone. Of the 1,316 remaining respondents, 102 broke off the survey (7.8\%), resulting in an analysis sample of N=1,214.

\textbf{Sample 4.} The fourth sample comes from Wave V.2 of the PUMA (Plattform für Umfragen, Methoden und empirische Analysen) project conducted by Statistics Austria in spring 2018 \citep{PUMA2019}. A random sample of 1,500 Austrian residents aged 16-74 was drawn from the Austrian population register and invited to participate in a web survey. Of the 695 individuals who completed the survey, 63 did not own a smartphone, resulting in an analysis sample of N=632.

\textbf{Sample 5.} The fifth sample comes from a web survey conducted among members of a German nonprobability online panel in January 2020 \citep{Keusch2021}. Panel members were invited through a survey-router system to participate in the survey, and 3,350 individuals started the survey. Only smartphone owners were eligible for the study, and quotas for gender, age, and frequency of smartphone usage were employed. Of the panel members who started the survey, 669 were screened out because of the quotas, because they reported not owning a smartphone, or because they did not live in Germany. Of the 2,681 remaining respondents, 156 broke off the survey (5.8\%), resulting in an analysis sample of N=2,525.

Descriptive statistics for the five samples are shown in Table \ref{tab:table_A1} in the Appendix. With the exception of age, with Sample 2 containing a larger proportion of older adults than the other samples, the five samples are comparable in terms of sociodemographic characteristics.

\subsection{Latent class analysis}
To identify different types of smartphone users, we conduct a latent class analysis (LCA) which classifies individuals based on their similarity in response patterns on a set of categorical variables \citep{McCutcheon1987}. LCA has been used in a number of previous studies to construct typologies of Internet usage \citep{Elhai2018, Hamka2014, Herzing2019, Holmes2011, Sell2014, Yates2020}.

We include 14 variables in the latent class models which were collected in all five samples, including frequency of smartphone use, self-rated smartphone skills, and twelve smartphone activities (Table \ref{tab:table_B1} in the Appendix). Frequency of smartphone use is measured on a five-point rating scale collapsed to three categories (several times a day, every day, several times a week or less). Self-rated smartphone skills are measured on a five-point rating scale collapsed to three categories (Advanced (5), Intermediate (4), Beginner (1-3)). For these two variables, the categories were collapsed because the response distributions of the original categories were highly skewed. Finally, twelve activities carried out on the smartphone are measured with a series of yes-no questions, reflecting types of activities that were examined in previous research on Internet usage, including social interaction (post social media), reading and writing emails (email), entertainment (games, streaming), commercial activities (online purchase, online banking), and information seeking (browse websites, view social media). The list was complemented by activities that are specifically carried out on smartphones, including taking photos, using GPS/location-aware apps, installing new apps, and using Bluetooth to connect the smartphone to other devices. A small proportion of missing values on these variables ($<$ 2\%) were imputed with a chained-equations algorithm by using the R \texttt{mice} package version 3.13.0 \citep{VanBuuren2011}. The missing values could not be imputed for two respondents in Sample 1 and for one respondent in Sample 3, reducing the analysis sample to N=2,621 (Sample 1) and N=1,213 (Sample 3). Descriptive statistics for the variables included in the LCA are shown in Table A1 in the Appendix. 

To estimate the latent class models, we use the R \texttt{poLCA} package version 1.4.1 \citep{Linzer2011}. We vary the number of classes in the LCA from two to ten and compute model fit criteria to select the best fitting model, including the log likelihood (LL), the Akaike information criterion (AIC), and the Bayesian information criterion (BIC), with lower values indicating a better model fit \citep{Nylund2007}. We also report the size and percentage of the smallest class.

\subsection{Sociodemographic and smartphone-related correlates}
As correlates of smartphone usage types, we collected data in the five samples on sociodemographic and other smartphone-related characteristics (see Table \ref{tab:table_A1} in the Appendix for descriptive statistics). The sociodemographic characteristics include \textit{gender} (male vs. female), \textit{age} (18-29 years, 30-39 years, 40-49 years, 50-59 years, 60+ years), and \textit{educational attainment} (no high school degree, high school degree, college degree). The categories for age and educational attainment are slightly different for Sample 2 and Sample 4 (see note in Table \ref{tab:table_A1} in the Appendix for the different categorization). Smartphone-related characteristics include the \textit{operating system} of the device (iOS, Android, other operating system). The data preparation and analysis were conducted in R version 4.0.4 \citep{RCoreTeam2021}.

\section{Results}
\subsection{Which smartphone usage types can be identified among smartphone owners?}
We identify smartphone usage types among smartphone owners by using data from the most recent sample, Sample 5, collected in 2020. In RQ3, we will then compare the results with data from the earlier samples, Samples 1-4, collected in 2016-2018. Varying the number of classes from two to ten shows that the BIC reaches a minimum at the six-class model, with a log likelihood and AIC that do not decrease substantially as more classes are included in the model (Table \ref{tab:table_1}). The six-class model also results in classes with a reasonable size, with the smallest class containing 156 individuals, 6\% of the overall sample. We therefore select the six-class solution for our analysis.

\begin{table}[H]
\caption{\label{tab:table_1} Model fit and diagnostic criteria for two to ten classes of smartphone usage (Sample 5).}
\centering
\begin{tabular}{lrrrrr}
\hline
\multicolumn{1}{c}{} & \multicolumn{3}{c}{Model fit criteria} & \multicolumn{2}{c}{Diagnostic criteria} \\
\hline
\# Classes & \multicolumn{1}{c}{LL} & \multicolumn{1}{c}{AIC} & \multicolumn{1}{c}{BIC} & \multicolumn{1}{c}{\begin{tabular}[c]{@{}c@{}}Smallest \\ class \\ (n)\end{tabular}} & \multicolumn{1}{c}{\begin{tabular}[c]{@{}c@{}}Smalles \\ class \\ (\%)\end{tabular}} \\
\hline
2 & -17,194.68 & 34,455.36 & 34,647.88 & 813 & 32 \\
3 & -16,716.83 & 33,533.66 & 33,825.36 & 289 & 11 \\
4 & -16,515.94 & 33,165.88 & 33,556.76 & 306 & 12 \\
5 & -16,406.31 & 32,980.62 & 33,470.67 & 134 & 5 \\
6 & \textbf{-16,317.85} & \textbf{32,837.70} & \textbf{33,426.93} & \textbf{156} & \textbf{6} \\
7 & -16,277.62 & 32,791.24 & 33,479.65 & 80 & 3 \\
8 & -16,249.80 & 32,769.60 & 33,557.19 & 82 & 3 \\
9 & -16,228.25 & 32,760.50 & 33,647.26 & 32 & 1 \\
10 & -16,202.97 & 32,743.94 & 33,729.89 & 26 & 1 \\
\hline
\multicolumn{6}{l}{\footnotesize Note. N=2,525. LL=log likelihood; AIC=Akaike information criterion; BIC=Bayesian} \\
\multicolumn{6}{l}{\footnotesize information criterion.}
\end{tabular}
\end{table}

We next examine the composition of the latent classes, with Table \ref{tab:table_2} showing the predictor variables (frequency of use, skills, types of activities) by smartphone usage class. We describe the six usage types as follows:

\textit{Advanced users} use their smartphone several times a day (82\%), rate their smartphone skills as advanced (59\%), and use their smartphone for each of the twelve activities (used by at least 87\%). They constitute the largest usage group, with almost half of the sample (44\%) categorized as advanced users.

\textit{Broad non-social-media users} use their smartphone several times a day (78\%), and mostly rate their smartphone skills as intermediate (39\%) or advanced (35\%), with an additional 25\% rating their skills as beginner. They use their smartphone for a large variety of activities (used by at least 58\%) with the exception of social media, with only 7\% posting content and 23\% viewing content on social media. They constitute 9\% of the sample.

\textit{Broad non-commercial users} use their smartphone several times a day (67\%) or every day (27\%), and rate their smartphone skills as intermediate (40\%), advanced (30\%), or beginner (30\%). They use their smartphone for a large variety of activities (used by at least 59\%) with the exception of commercial activities, with only 37\% using their smartphone for online banking and 45\% for online purchases. They constitute the second largest usage group, with almost one fourth of the sample (23\%) categorized as broad non-commercial users.

\textit{Basic general users} use their smartphone several times a day (31\%), every day (35\%), or several times a week or less (35\%), and mostly rate their smartphone skills as beginner (47\%) or intermediate (39\%). Almost all of them use their smartphone to take photos (91\%) and browse websites (90\%). Other popular activities include using GPS/location-aware apps (83\%), installing new apps (80\%), and reading and/or writing emails (79\%). They constitute 11\% of the sample.

\textit{Social media and information users} use their smartphone several times a day (45\%), with an additional 29\% using their smartphone every day and 26\% several times a week or less, and mostly rate their smartphone skills as beginner (62\%) or intermediate (30\%). They mainly use their smartphone to browse websites (76\%) and use social media, with 74\% viewing content and 72\% posting content on social media. They constitute 6\% of the sample.

\textit{Camera users} mostly use their smartphone several times a week or less (62\%), with only 23\% using their smartphone every day and 15\% several times a day, and rate their smartphone skills as beginner (82\%). They mainly use their smartphone to take photos (77\%). Other popular activities include browsing websites (48\%), and reading and/or writing emails (45\%). Only a small share of users ($\leq$ 13\%) are engaged with the other nine activities. They constitute 7\% of the sample.

\begin{landscape}
\begin{center}
\begin{longtable}{lrrrrrr}
\caption{\label{tab:table_2} Predictor variables by class of smartphone usage (Sample 5).} \\
\hline
\multicolumn{1}{c}{Variables} & \multicolumn{1}{c}{\begin{tabular}[t]{@{}c@{}}Advanced \\ users\end{tabular}} & \multicolumn{1}{c}{\begin{tabular}[t]{@{}c@{}}Broad non-\\ social-media\\ users\end{tabular}} & \multicolumn{1}{c}{\begin{tabular}[t]{@{}c@{}}Broad non-\\ commercial\\ users\end{tabular}} & \multicolumn{1}{c}{\begin{tabular}[t]{@{}c@{}}Basic \\ general \\ users\end{tabular}} & \multicolumn{1}{c}{\begin{tabular}[t]{@{}c@{}}Social media \\ and information \\ users\end{tabular}} & \multicolumn{1}{c}{\begin{tabular}[t]{@{}c@{}}Camera\\ users\end{tabular}} \\
 & \multicolumn{1}{c}{\%} & \multicolumn{1}{c}{\%} & \multicolumn{1}{c}{\%} & \multicolumn{1}{c}{\%} & \multicolumn{1}{c}{\%} & \multicolumn{1}{c}{\%} \\
\hline
\endfirsthead
\caption{Predictor variables by class of smartphone usage (Sample 5) (\textit{continued}).} \\
\hline
\multicolumn{1}{c}{Variables} & \multicolumn{1}{c}{\begin{tabular}[t]{@{}c@{}}Advanced \\ users\end{tabular}} & \multicolumn{1}{c}{\begin{tabular}[t]{@{}c@{}}Broad non-\\ social-media\\ users\end{tabular}} & \multicolumn{1}{c}{\begin{tabular}[t]{@{}c@{}}Broad non-\\ commercial\\ users\end{tabular}} & \multicolumn{1}{c}{\begin{tabular}[t]{@{}c@{}}Basic \\ general \\ users\end{tabular}} & \multicolumn{1}{c}{\begin{tabular}[t]{@{}c@{}}Social media \\ and information \\ users\end{tabular}} & \multicolumn{1}{c}{\begin{tabular}[t]{@{}c@{}}Camera\\ users\end{tabular}} \\
 & \multicolumn{1}{c}{\%} & \multicolumn{1}{c}{\%} & \multicolumn{1}{c}{\%} & \multicolumn{1}{c}{\%} & \multicolumn{1}{c}{\%} & \multicolumn{1}{c}{\%} \\
\hline
\endhead
Class size & 44 & 9 & 23 & 11 & 6 & 7 \\
\hline
Frequency of smartphone use &  &  &  &  &  &  \\
\hspace{3mm} Several times a day & 82 & 78 & 67 & 31 & 45 & 15 \\
\hspace{3mm} Every day & 13 & 19 & 27 & 35 & 29 & 23 \\
\hspace{3mm} Several times a week or less & 5 & 3 & 6 & 35 & 26 & 62 \\
Smartphone skills &  &  &  &  &  &  \\
\hspace{3mm} Advanced (5) & 59 & 35 & 30 & 14 & 8 & 7 \\
\hspace{3mm} Intermediate (4) & 32 & 39 & 40 & 39 & 30 & 11 \\
\hspace{3mm} Beginner (1-3) & 9 & 25 & 30 & 47 & 62 & 82 \\
Smartphone activities &  &  &  &  &  &  \\
\hspace{3mm} Browse websites & 100 & 100 & 99 & 90 & 76 & 48 \\
\hspace{3mm} Email & 99 & 100 & 92 & 79 & 65 & 45 \\
\hspace{3mm} Photo & 100 & 97 & 99 & 91 & 64 & 77 \\
\hspace{3mm} View social media & 100 & 23 & 100 & 11 & 74 & 11 \\
\hspace{3mm} Post social media & 93 & 7 & 70 & 0 & 72 & 0 \\
\hspace{3mm} Online purchase & 100 & 96 & 45 & 20 & 40 & 2 \\
\hspace{3mm} Online banking & 94 & 85 & 37 & 28 & 38 & 3 \\
\hspace{3mm} Install apps & 100 & 97 & 90 & 80 & 31 & 5 \\
\hspace{3mm} GPS & 99 & 91 & 85 & 83 & 42 & 11 \\
\hspace{3mm} Bluetooth & 94 & 73 & 59 & 41 & 23 & 7 \\
\hspace{3mm} Games & 87 & 58 & 61 & 34 & 49 & 13 \\
\hspace{3mm} Streaming & 99 & 76 & 78 & 38 & 42 & 5 \\
\hline
N & 1,103 & 226 & 583 & 289 & 156 & 168 \\
\hline
\end{longtable}
\end{center}
\end{landscape}

\subsection{How do the smartphone usage types differ in sociodemographic and smartphone-related characteristics?}
We next examine how the smartphone usage types differ in sociodemographic and smartphone-related characteristics, with Table \ref{tab:table_3} showing the characteristics by smartphone usage class. To test whether differences are statistically significant, we conduct Pearson’s Chi-squared tests and report the p-values. We find significant differences between the smartphone usage types by gender (p=0.005). \textit{Broad non-social-media users} and \textit{basic general users} have a larger proportion of men than women (both 58\% vs. 42\%) whereas the other usage types are relatively balanced with regard to gender.

The results also indicate that the smartphone usage classes have a significantly different age composition (p$<$0.001). The large majority of \textit{advanced users} belong to the younger age groups, with two thirds of users aged between 18 and 39. For \textit{broad non-social-media users} and \textit{broad non-commercial users}, the age composition is somewhat older, but still a majority is below the age of 40 in these groups. Finally, the usage types reflecting more narrow and less frequent smartphone use, \textit{camera users} and \textit{basic general users}, constitute the oldest groups of smartphone users, with more than half of users aged 50 and older (59\% and 52\%, respectively) and an additional 20\% and 28\% aged between 40 and 49. An exception are \textit{social media and information users} who, again, tend to be younger, with 52\% of users below the age of 40.

The smartphone usage types also differ significantly by educational attainment (p$<$ 0.001). \textit{Advanced users} and \textit{broad non-social-media users} have the highest proportion of college graduates (32\% and 29\%) and high school graduates (30\% and 28\%) among the smartphone usage types. \textit{Social media and information users} have similar levels of education, with 28\% having a college degree and 26\% having a high school degree. The other three usage types, in turn, comprise individuals with lower levels of educational attainment, with \textit{camera users} having the highest proportion of people without a high school degree (57\%).

We also find significant differences between the smartphone usage types by operating system (p$<$0.001). While the majority of people across all usage types report owning an Android smartphone, \textit{advanced users} and \textit{social media and information users} are more likely to own an iPhone (33\% and 28\%) than \textit{broad non-social-media users} (21\%) and \textit{broad non-commercial users} (19\%) who, in turn, are more likely to be iPhone owners than \textit{basic general users} (16\%) and \textit{camera users} (12\%).

\begin{landscape}
\begin{center}
\begin{longtable}{lrrrrrrr}
\caption{\label{tab:table_3} Sociodemographic and smartphone-related characteristics by class of smartphone usage (Sample 5).} \\
\hline
\multicolumn{1}{c}{Variables} & \multicolumn{1}{c}{\begin{tabular}[t]{@{}c@{}}Advanced \\ users\end{tabular}} & \multicolumn{1}{c}{\begin{tabular}[t]{@{}c@{}}Broad non-\\ social-media\\ users\end{tabular}} & \multicolumn{1}{c}{\begin{tabular}[t]{@{}c@{}}Broad non-\\ commercial\\ users\end{tabular}} & \multicolumn{1}{c}{\begin{tabular}[t]{@{}c@{}}Basic \\ general \\ users\end{tabular}} & \multicolumn{1}{c}{\begin{tabular}[t]{@{}c@{}}Social media \\ and information \\ users\end{tabular}} & \multicolumn{1}{c}{\begin{tabular}[t]{@{}c@{}}Camera\\ users\end{tabular}} & \multicolumn{1}{l}{p-Value} \\
 & \multicolumn{1}{c}{\%} & \multicolumn{1}{c}{\%} & \multicolumn{1}{c}{\%} & \multicolumn{1}{c}{\%} & \multicolumn{1}{c}{\%} & \multicolumn{1}{c}{\%} & \multicolumn{1}{l}{} \\
\hline
\endhead
Gender &  &  &  &  &  &  & 0.005 \\
\hspace{3mm} Female & 51 & 42 & 52 & 42 & 49 & 53 &  \\
\hspace{3mm} Male & 49 & 58 & 48 & 58 & 51 & 47 &  \\
Age &  &  &  &  &  &  & \textless{}0.001 \\
\hspace{3mm} 18-29 & 38 & 16 & 26 & 8 & 31 & 8 &  \\
\hspace{3mm} 30-39 & 28 & 23 & 23 & 13 & 21 & 14 &  \\
\hspace{3mm} 40-49 & 20 & 31 & 27 & 28 & 22 & 20 &  \\
\hspace{3mm} 50-59 & 12 & 21 & 18 & 36 & 19 & 39 &  \\
\hspace{3mm} 60+ & 2 & 9 & 6 & 16 & 8 & 20 &  \\
Educational attainment &  &  &  &  &  &  & \textless{}0.001 \\
\hspace{3mm} No high school degree & 38 & 43 & 50 & 51 & 46 & 57 &  \\
\hspace{3mm} High school degree & 30 & 28 & 24 & 19 & 26 & 18 &  \\
\hspace{3mm} College degree & 32 & 29 & 25 & 29 & 28 & 24 &  \\
Operating system &  &  &  &  &  &  & \textless{}0.001 \\
\hspace{3mm} iOS & 33 & 21 & 19 & 16 & 28 & 12 &  \\
\hspace{3mm} Android & 64 & 78 & 76 & 79 & 63 & 81 &  \\
\hspace{3mm} Other operating system & 3 & 0 & 4 & 6 & 10 & 7 & \\
\hline
\end{longtable}
\end{center}
\end{landscape}

\subsection{How stable are the smartphone usage types across time and samples?}
Finally, we examine whether the smartphone usage types have changed between 2016 and 2020 by comparing the results presented for Sample 5 with the four earlier samples, collected in 2016 (Sample 1), 2017 (Sample 2 and 3), and 2018 (Sample 4). For each of these samples, we repeated the LCA, varying the number of classes from two to ten. 

In both Sample 1 and Sample 2, the results show that the BIC reaches a minimum at the six-class model, providing support for a six-class solution in line with Sample 5 (Table \ref{tab:table_C1} in the Appendix). When examining the composition of these latent classes, the same six smartphone usage types can be identified, including \textit{advanced users}, \textit{broad non-social-media users}, \textit{broad non-commercial users}, \textit{basic general users}, \textit{social media and information users}, and \textit{camera users} (Tables \ref{tab:table_D1} and \ref{tab:table_E1} in the Appendix). While the size of the latent classes is similar across Sample 1 and Sample 5, the class size is notably different in Sample 2: There is a considerably smaller proportion of \textit{advanced users} in Sample 2 compared to Sample 5 (17\% vs. 44\%) but a larger proportion of \textit{basic general users} (25\% vs. 11\%), \textit{camera users} (20\% vs. 7\%), and \textit{social media and information users} (12\% vs. 6\%).

The LCA results for Sample 3 and Sample 4, however, are in favor of a four-class solution, with the BIC reaching a minimum at the four-class model. A subset of the smartphone usage types can be identified in these samples.

In Sample 3, we can identify \textit{advanced users} (47\% of the sample; Table \ref{tab:table_F1} in the Appendix), \textit{broad non-commercial users} (26\% of the sample), \textit{basic general users} (13\% of the sample), and \textit{camera users} (14\% of the sample). Similarly, we can identify \textit{advanced users} (39\% of the sample; Table \ref{tab:table_G1} in the Appendix) and \textit{camera users} (23\% of the sample) in Sample 4, and additionally \textit{broad non-social media users} and \textit{social media and information users} (each 19\% of the sample).

Overall, the composition of these latent classes as well as their sociodemographic and smartphone-related characteristics are robust across samples, but there are a few differences (see Table \ref{tab:table_D1}-Table \ref{tab:table_G2} in the Appendix):

\textit{Advanced users} in the earlier samples are less likely to engage in certain types of activities on their smartphone, although they still use their smartphone for a larger variety of activities compared to the other types of smartphone users. For example, advanced users in Sample 2, compared to those in Sample 5, are less likely to use Bluetooth (77\% vs. 94\%), use online banking (70\% vs. 94\%) or play games on their smartphone (60\% vs. 87\%). In addition, there is a slightly larger proportion of men among advanced users in Sample 2 (56\%) and Sample 4 (54\%) than in Sample 5 (49\%).

\textit{Broad non-social-media users} have a slightly larger proportion of men in Sample 1 (63\%) and Sample 2 (66\%) than in Sample 5 (58\%), but a slightly smaller proportion of men in Sample 4 (50\%).

\textit{Broad non-commercial users} in Sample 2 are considerably less likely to use their smartphone for a number of activities compared to Sample 5, including streaming videos or music on their smartphone (48\% vs. 78\%), playing games (33\% vs. 61\%), installing apps (62\% vs. 90\%), posting content on social media (46\% vs. 70\%), and using Bluetooth (42\% vs. 59\%).

\textit{Basic general users} in Sample 2 are considerably less likely to install apps on their smartphone (30\%) than those in Sample 5 (80\%).

\textit{Social media and information users} in Sample 2 are more likely to take photos on their smartphone compared to Sample 5 (84\% vs. 64\%). In addition, they have a larger proportion of women in Sample 1 (57\%), Sample 2 (57\%), and Sample 4 (58\%) than in Sample 5 (49\%). They also have an older age distribution in Sample 1, with 36\% below the age of 40, and Sample 2, with 16\% below the age of 38, compared to Sample 5 where more than half of the users are younger than 40. There is a larger proportion of people without a high school degree in Sample 1 (62\%), Sample 2 (57\%), and Sample 4 (60\%) than in Sample 5 (46\%). Finally, a smaller proportion of social media and information users own an iPhone in Sample 1 compared to Sample 5 (14\% vs. 28\%).

\textit{Camera users} have a larger proportion of men in Sample 1 (55\%) than in Sample 5 (47\%), but a smaller proportion of men in Sample 2 (43\%). There is also a larger proportion of people without a high school degree in Sample 1 (70\%) and Sample 2 (67\%) than in Sample 5 (57\%). In addition, camera users in Sample 2 are even older than in Sample 5 (53\% aged 58+ vs. 20\% aged 60+).

\section{Discussion}
Over the last decade, smartphones have become relevant for an increasing number of activities in peoples’ daily lives, from communication to health-related and commercial activities. Although a growing number of people in the general population have access to these devices, a substantial proportion of those still lack the necessary skills and experience to fully use and take advantage of their smartphone.

In this paper, we aim to contribute to the small body of research on the second-level smartphone divide by constructing a typology of smartphone users based on survey data from five samples of smartphone owners collected in Germany and Austria between 2016 and 2020. We classify individuals based on their frequency of smartphone use, self-rated smartphone skills, and activities carried out on their smartphone, and identify six distinct types of smartphone users, with a subset of these types found across all samples. 

Three of these usage types represent smartphone owners who use their device frequently and mostly rate their smartphone skills as advanced or intermediate, but differ in the types of activities that they carry out on their smartphone: \textit{Advanced users} use their smartphone for the full set of activities that we examined in the study whereas \textit{broad non-social-media users} use their device for all activities but viewing and posting content on social media. Similarly, \textit{broad non-commercial users} use their smartphone for a large variety of activities with the exception of online banking and making online purchases. Advanced users have also been identified in previous typologies (e.g., \citeauthor{DeCanio2016}, 2016: “supersmartphoners”; \citeauthor{Reisdorf2017}, 2017: “broad users”; \citeauthor{Yates2020}, 2020: “extensive users”) as well as broad non-social-media users (e.g., \citeauthor{Yates2020}, 2020: “general (no social media) users”). Smartphone owners who belong to one of these usage types tend to be younger and have higher levels of educational attainment, which is in line with previous research in this area \citep{Andone2016, Fortunati2014, Serrano-Cinca2018}. Advanced users also have the highest proportion of iPhone owners, which is consistent with recent research showing that iPhone users tend to be younger and have higher levels of educational attainment \citep{Keusch2020}.

The other three usage types, in turn, represent smartphone owners who use their device less frequently, mostly rate their smartphone skills as beginner or intermediate, and use their device for a limited number of activities: \textit{Camera users} mainly use their smartphone to take photos whereas \textit{social media and information users} mostly use their device to browse websites and use social media. Finally, \textit{basic general users} use their smartphone to take photos, browse websites, use GPS/location-aware apps, install new apps, and read and/or write emails. Social media and information users have also been identified in previous typologies (e.g., \citeauthor{Yates2020}, 2020: “social and entertainment media only users”) as well as basic general users (e.g., \citeauthor{Yates2020}, 2020: “utility users”). Smartphone owners who belong to one of these usage types tend to be older, with the exception of social media and information users, and have lower levels of educational attainment, consistent with prior research \citep{Andone2016, Fortunati2014, Serrano-Cinca2018}. Overall, these results are robust across the five samples collected between 2016 and 2020.

A key take-away message from our study is that, although smartphones have become deeply integrated into most people’s daily life, smartphone usage patterns continue to be highly diverse and a sizable proportion of smartphone owners still use their device for a rather narrow set of activities. With regard to the Internet, \citet[p. 7]{Selwyn2005} noted that it is “not one technology but means different things to different people and is used in different ways for different purposes”. The present paper shows that the same can be said about smartphone technology almost two decades later. As smartphones become increasingly relevant in our daily life, additional support needs to be provided to users with lower levels of skills and experience. For example, these users could be assisted with installing and using apps on their smartphone that are required to access certain types of services, such as medical apps to prevent and treat chronic diseases. A better understanding of prevalent usage patterns and their association with sociodemographic and smartphone-related characteristics might be a first step in this direction.

Our study is not free of limitations. First, we relied on self-reported measures of smartphone usage that are likely to be affected by recall error: Respondents may not accurately remember the frequency, duration, or variety of their smartphone usage and the self-reports are, thus, likely to be biased \citep{Boase2013}. Future research might consider relying on passively collected measures of smartphone usage \citep{Festic2021, Hamka2014}. Second, while we aimed to cover the most popular activities carried out on smartphones, our list of activities may not be exhaustive and it is likely that we omitted additional key activities. Third, while we investigated the robustness of our smartphone usage typology across multiple samples, we were only able to examine a relatively short period of four years. Future research might consider studying changes in the second-level smartphone divide over a longer period of time.

\bibliography{references}

\newpage
\appendix

\begin{landscape}
\section{Sample composition}
\vspace{-8mm}
\setcounter{table}{0} \renewcommand{\thetable}{A.\arabic{table}}
\begin{center}
\begin{ThreePartTable}
\begin{TableNotes}
    \item[a] The age categories are different for Sample 2 (18-27, 28-37, 38-47, 48-57, 58+) and Sample 4 (18-29, 30-49, 50+).
    \item[b] The categories for educational attainment are different for Sample 4 (no high school degree, high school degree).
\end{TableNotes}
\begin{longtable}{lrrrrrrrrrr}
\caption{\label{tab:table_A1} Descriptive statistics.} \\
\hline
 & \multicolumn{2}{c}{\begin{tabular}[c]{@{}c@{}}Sample 1\\ 2016\end{tabular}} & \multicolumn{2}{c}{\begin{tabular}[c]{@{}c@{}}Sample 2\\ 2017\end{tabular}} & \multicolumn{2}{c}{\begin{tabular}[c]{@{}c@{}}Sample 3\\ 2017\end{tabular}} & \multicolumn{2}{c}{\begin{tabular}[c]{@{}c@{}}Sample 4\\ 2018\end{tabular}} & \multicolumn{2}{c}{\begin{tabular}[c]{@{}c@{}}Sample 5\\ 2020\end{tabular}} \\
\hline
 & \multicolumn{1}{c}{\%} & \multicolumn{1}{c}{Missing} & \multicolumn{1}{c}{\%} & \multicolumn{1}{c}{Missing} & \multicolumn{1}{c}{\%} & \multicolumn{1}{c}{Missing} & \multicolumn{1}{c}{\%} & \multicolumn{1}{c}{Missing} & \multicolumn{1}{c}{\%} & \multicolumn{1}{c}{Missing} \\
\hline
\endfirsthead
\caption{Descriptive statistics (\textit{continued}).} \\
\hline
 & \multicolumn{2}{c}{\begin{tabular}[c]{@{}c@{}}Sample 1\\ 2016\end{tabular}} & \multicolumn{2}{c}{\begin{tabular}[c]{@{}c@{}}Sample 2\\ 2017\end{tabular}} & \multicolumn{2}{c}{\begin{tabular}[c]{@{}c@{}}Sample 3\\ 2017\end{tabular}} & \multicolumn{2}{c}{\begin{tabular}[c]{@{}c@{}}Sample 4\\ 2018\end{tabular}} & \multicolumn{2}{c}{\begin{tabular}[c]{@{}c@{}}Sample 5\\ 2020\end{tabular}} \\
\hline
 & \multicolumn{1}{c}{\%} & \multicolumn{1}{c}{Missing} & \multicolumn{1}{c}{\%} & \multicolumn{1}{c}{Missing} & \multicolumn{1}{c}{\%} & \multicolumn{1}{c}{Missing} & \multicolumn{1}{c}{\%} & \multicolumn{1}{c}{Missing} & \multicolumn{1}{c}{\%} & \multicolumn{1}{c}{Missing} \\
\hline
\endhead
Gender &  & 0 &  & 1 &  & 0 &  & 0 &  & 0 \\
\hspace{3mm} Female & 49.8 &  & 49.2 &  & 50.0 &  & 51.1 &  & 49.6 &  \\
\hspace{3mm} Male & 50.2 &  & 50.8 &  & 50.0 &  & 48.9 &  & 50.4 &  \\
Age\tnote{a} &  & 0 &  & 1 &  & 0 &  & 2 &  & 0 \\
\hspace{3mm} 18-29 & 28.9 &  & 10.8 &  & 29.2 &  & 29.7 &  & 27.3 &  \\
\hspace{3mm} 30-39 & 23.0 &  & 18.1 &  & 23.1 &  & 36.8 &  & 23.4 &  \\
\hspace{3mm} 40-49 & 24.5 &  & 17.6 &  & 24.3 &  & -- &  & 23.4 &  \\
\hspace{3mm} 50-59 & 17.5 &  & 26.0 &  & 17.4 &  & 33.5 &  & 19.1 &  \\
\hspace{3mm} 60+ & 6.1 &  & 27.5 &  & 6.0 &  & -- &  & 6.7 &  \\
Educational attainment\tnote{b} &  & 10 &  & 1 &  & 1 &  & 0 &  & 0 \\
\hspace{3mm} No high school degree & 47.4 &  & 46.0 &  & 48.4 &  & 51.6 &  & 44.6 &  \\
\hspace{3mm} High school degree & 28.3 &  & 22.6 &  & 28.0 &  & 48.4 &  & 26.2 &  \\
\hspace{3mm} College degree & 24.3 &  & 31.4 &  & 23.7 &  & -- &  & 29.1 &  \\
Operating system &  & 20 &  & -- &  & 10 &  & -- &  & 11 \\
\hspace{3mm} iOS & 22.4 &  & -- &  & 24.4 &  & -- &  & 25.0 &  \\
\hspace{3mm} Android & 70.6 &  & -- &  & 68.4 &  & -- &  & 70.8 &  \\
\hspace{3mm} Other operating system & 7.0 &  & -- &  & 7.1 &  & -- &  & 4.2 &  \\
Frequency of smartphone use &  & 0 &  & 0 &  & 0 &  & 0 &  & 0 \\
\hspace{3mm} Several times a day & 77.0 &  & 69.3 &  & 70.2 &  & 68.8 &  & 65.6 &  \\
\hspace{3mm} Every day & 13.4 &  & 14.4 &  & 14.7 &  & 16.6 &  & 20.8 &  \\
\hspace{3mm} Several times a week or less & 9.7 &  & 16.4 &  & 15.1 &  & 14.6 &  & 13.6 &  \\
Smartphone skills &  & 0 &  & 0 &  & 0 &  & 0 &  & 0 \\
\hspace{3mm} Advanced (5) & 35.5 &  & 23.9 &  & 41.1 &  & 33.2 &  & 38.3 &  \\
\hspace{3mm} Intermediate (4) & 35.5 &  & 28.5 &  & 31.8 &  & 27.2 &  & 34.0 &  \\
\hspace{3mm} Beginner (1-3) & 29.0 &  & 47.6 &  & 27.1 &  & 39.6 &  & 27.7 &  \\
Smartphone activities &  &  &  &  &  &  &  &  &  &  \\
\hspace{3mm} Browse websites & 90.5 & 0 & 76.7 & 0 & 91.3 & 0 & 87.0 & 0 & 93.5 & 0 \\
\hspace{3mm} Email & 85.8 & 0 & 76.2 & 0 & 87.3 & 0 & 79.7 & 0 & 89.5 & 0 \\
\hspace{3mm} Photo & 96.6 & 0 & 90.1 & 0 & 96.5 & 0 & 95.1 & 0 & 94.7 & 0 \\
\hspace{3mm} View social media & 75.6 & 0 & 46.8 & 0 & 75.8 & 0 & 66.6 & 0 & 75.4 & 0 \\
\hspace{3mm} Post social media & 61.5 & 0 & 26.2 & 0 & 61.2 & 0 & 44.8 & 0 & 61.8 & 0 \\
\hspace{3mm} Online purchase & 51.7 & 0 & 33.1 & 0 & 61.8 & 0 & 42.2 & 0 & 67.5 & 0 \\
\hspace{3mm} Online banking & 45.1 & 0 & 25.3 & 0 & 50.5 & 0 & 43.7 & 0 & 62.8 & 0 \\
\hspace{3mm} Install apps & 83.4 & 0 & 42.3 & 0 & 81.7 & 0 & 66.1 & 0 & 84.5 & 0 \\
\hspace{3mm} GPS & 83.0 & 0 & 58.5 & 0 & 80.0 & 0 & 75.5 & 0 & 83.8 & 0 \\
\hspace{3mm} Bluetooth & 47.7 & 0 & 33.4 & 0 & 53.0 & 0 & 47.0 & 0 & 67.6 & 0 \\
\hspace{3mm} Games & 62.6 & 0 & 29.4 & 0 & 65.2 & 0 & 34.8 & 0 & 65.0 & 0 \\
\hspace{3mm} Streaming & 68.3 & 0 & 39.0 & 0 & 72.0 & 0 & 65.8 & 0 & 75.4 & 0 \\
\hline
N & \multicolumn{2}{c}{2,621} & \multicolumn{2}{c}{2,186} & \multicolumn{2}{c}{1,213} & \multicolumn{2}{c}{632} & \multicolumn{2}{c}{2,525}\\
\hline
\insertTableNotes
\end{longtable}
\end{ThreePartTable}
\end{center}
\end{landscape}

\section{Questionnaire}
\vspace{-8mm}
\setcounter{table}{0} \renewcommand{\thetable}{B.\arabic{table}}
\begin{center}
\begin{longtable}{lll}
\caption{\label{tab:table_B1} Questionnaire.}\\
\hline
Variable & Question text & Response options \\
\hline
\endfirsthead
\caption{\label{table_B1} Questionnaire (\textit{continued}).}\\
\hline
Variable & Question text & Response options \\
\hline
\endhead
\begin{tabular}[t]{@{}l@{}}Frequency of\\ smartphone use\end{tabular} & \begin{tabular}[t]{@{}l@{}}How often do you use a smartphone for activities\\ other than phone calls or text messaging?\end{tabular} & \begin{tabular}[t]{@{}l@{}}Several times a day\\ Every day\\ Several times a week\\ Several times a month\\ Once a month or less\end{tabular} \\
\begin{tabular}[t]{@{}l@{}}Smartphone\\ skills\end{tabular} & \begin{tabular}[t]{@{}l@{}}Generally, how would you rate your skills of using\\ your smartphone?\end{tabular} & \begin{tabular}[t]{@{}l@{}}1 Beginner\\ 2\\ 3\\ 4\\ 5 Advanced\end{tabular} \\
 & \begin{tabular}[t]{@{}l@{}}Do you use your smartphone for the following\\ activities?\end{tabular} &  \\
\begin{tabular}[t]{@{}l@{}}Browse\\ websites\end{tabular} & Browsing websites & \begin{tabular}[t]{@{}l@{}}Yes\\ No\end{tabular} \\
Email & Reading and/or writing email & \begin{tabular}[t]{@{}l@{}}Yes\\ No\end{tabular} \\
Photo & Taking photos & \begin{tabular}[c]{@{}l@{}}Yes\\ No\end{tabular} \\
\begin{tabular}[t]{@{}l@{}}View social\\ media\end{tabular} & \begin{tabular}[t]{@{}l@{}} Looking at content on social media websites/apps\\ (for example looking at text, images, videos on\\ Facebook, Twitter, Instagram)\end{tabular} & \begin{tabular}[t]{@{}l@{}}Yes\\ No\end{tabular} \\
\begin{tabular}[t]{@{}l@{}}Post social\\ media\end{tabular} & \begin{tabular}[t]{@{}l@{}}Posting content to social media websites/apps\\ (for example posting text, images, videos on\\ Facebook, Twitter, Instagram)\end{tabular} & \begin{tabular}[t]{@{}l@{}}Yes\\ No\end{tabular} \\
\begin{tabular}[t]{@{}l@{}}Online\\ purchase\end{tabular} & \begin{tabular}[t]{@{}l@{}}Making purchases (for example buying books or\\ clothes, booking train tickets, ordering food)\end{tabular} & \begin{tabular}[t]{@{}l@{}}Yes\\ No\end{tabular} \\
\begin{tabular}[t]{@{}l@{}}Online\\ banking\end{tabular} & \begin{tabular}[t]{@{}l@{}}Online banking (for example checking account\\ balance, transferring money)\end{tabular} & \begin{tabular}[t]{@{}l@{}}Yes\\ No\end{tabular} \\
Install apps & \begin{tabular}[t]{@{}l@{}}Installing new apps (for example from iTunes,\\ Google Play Store)\end{tabular} & \begin{tabular}[t]{@{}l@{}}Yes\\ No\end{tabular} \\
GPS & \begin{tabular}[t]{@{}l@{}}Using GPS/location-aware apps (for example\\ Google Maps, Foursquare, Yelp)\end{tabular} & \begin{tabular}[t]{@{}l@{}}Yes\\ No\end{tabular} \\
Bluetooth & \begin{tabular}[t]{@{}l@{}}Connecting to other electronic devices via Bluetooth\\ (for example smart-watches, fitness bracelets,\\ step counter)\end{tabular} & \begin{tabular}[t]{@{}l@{}}Yes\\ No\end{tabular} \\
Games & Playing games & \begin{tabular}[t]{@{}l@{}}Yes\\ No\end{tabular} \\
Streaming & Streaming videos or music & \begin{tabular}[t]{@{}l@{}}Yes\\ No\end{tabular} \\
\hline
\end{longtable}
\end{center}

\begin{landscape}
\section{Latent class analysis (Sample 1-Sample 4)}
\vspace{-8mm}
\setcounter{table}{0} \renewcommand{\thetable}{C.\arabic{table}}
\begin{table}[H]
\scriptsize
\caption{\label{tab:table_C1} Goodness of fit for two to ten classes of smartphone usage.}
\centering
\begin{tabular}{lrrrrrrrrrrrr}
\hline
 & \multicolumn{3}{c}{\begin{tabular}[c]{@{}c@{}}Sample 1\\ 2016\end{tabular}} & \multicolumn{3}{c}{\begin{tabular}[c]{@{}c@{}}Sample 2\\ 2017\end{tabular}} & \multicolumn{3}{c}{\begin{tabular}[c]{@{}c@{}}Sample 3\\ 2017\end{tabular}} & \multicolumn{3}{c}{\begin{tabular}[c]{@{}c@{}}Sample 4\\ 2018\end{tabular}} \\
\hline
\multicolumn{1}{c}{\# Classes} & \multicolumn{1}{c}{LL} & \multicolumn{1}{c}{AIC} & \multicolumn{1}{c}{BIC} & \multicolumn{1}{c}{LL} & \multicolumn{1}{c}{AIC} & \multicolumn{1}{c}{BIC} & \multicolumn{1}{c}{LL} & \multicolumn{1}{c}{AIC} & \multicolumn{1}{c}{BIC} & \multicolumn{1}{c}{LL} & \multicolumn{1}{c}{AIC} & \multicolumn{1}{c}{BIC} \\
\hline
2 & -18,355.59 & 36,777.18 & 36,970.94 & -16,942.82 & 33,951.64 & 34,139.40 & -8,432.766 & 16,931.53 & 17,099.86 & -4,984.907 & 10,035.814 & 10,182.63 \\
3 & -18,007.46 & 36,114.91 & 36,408.48 & -16,490.92 & 33,081.84 & 33,366.33 & -8,158.225 & 16,416.45 & 16,671.49 & -4,899.151 & 9,898.301 & 10,120.75 \\
4 & -17,594.93 & 35,323.86 & 35,717.24 & -16,378.65 & 32,891.31 & 33,272.52 & \textbf{-8,049.971} & \textbf{16,233.94} & \textbf{16,575.70} & \textbf{-4,815.937} & \textbf{9,765.874} & \textbf{10,063.95} \\
5 & -17,503.02 & 35,174.04 & 35,667.23 & -16,244.41 & 32,656.82 & 33,134.77 & -8,010.048 & 16,188.10 & 16,616.57 & -4,771.088 & 9,710.176 & 10,083.88 \\
6 & \textbf{-17,431.42} & \textbf{35,064.84} & \textbf{35,657.84} & \textbf{-16,166.67} & \textbf{32,535.35} & \textbf{33,110.02} & -7,983.129 & 16,168.26 & 16,683.44 & -4,735.303 & 9,672.606 & 10,121.94 \\
7 & -17,386.15 & 35,008.30 & 35,701.11 & -16,153.10 & 32,542.21 & 33,213.61 & -7,944.098 & 16,124.20 & 16,726.10 & -4,708.222 & 9,652.444 & 10,177.41 \\
8 & -17,351.09 & 34,972.18 & 35,764.80 & -16,103.31 & 32,476.62 & 33,244.75 & -7,920.505 & 16,111.01 & 16,799.62 & -4,681.096 & 9,632.192 & 10,232.79 \\
9 & -17,326.88 & 34,957.76 & 35,850.20 & -16,065.45 & 32,434.90 & 33,299.75 & -7,896.315 & 16,096.63 & 16,871.96 & -4,660.039 & 9,624.078 & 10,300.31 \\
10 & -17,309.98 & 34,957.96 & 35,950.21 & -16,043.34 & 32,424.68 & 33,386.27 & -7,887.485 & 16,112.97 & 16,975.01 & -4,642.332 & 9,622.665 & 10,374.53 \\
\hline
N & \multicolumn{3}{c}{2,621} & \multicolumn{3}{c}{2,186} & \multicolumn{3}{c}{1,213} & \multicolumn{3}{c}{632} \\
\hline
\multicolumn{13}{l}{Note. LL=log likelihood; AIC=Akaike information criterion; BIC=Bayesian information criterion.}
\end{tabular}
\end{table}

\newpage
\section{Sample 1}
\vspace{-8mm}
\setcounter{table}{0} \renewcommand{\thetable}{D.\arabic{table}}
\begin{center}
\begin{longtable}{lrrrrrr}
\caption{\label{tab:table_D1} Predictor variables by class of smartphone usage.} \\
\hline
Variables & \multicolumn{1}{c}{\begin{tabular}[t]{@{}c@{}}Advanced\\ users\end{tabular}} & \multicolumn{1}{c}{\begin{tabular}[t]{@{}c@{}}Broad non-\\ social-media \\ users\end{tabular}} & \multicolumn{1}{c}{\begin{tabular}[t]{@{}c@{}}Broad non-\\ commercial\\ users\end{tabular}} & \multicolumn{1}{c}{\begin{tabular}[t]{@{}c@{}}Basic\\ general\\ users\end{tabular}} & \multicolumn{1}{c}{\begin{tabular}[t]{@{}c@{}}Social media\\ and information\\ users\end{tabular}} & \multicolumn{1}{c}{\begin{tabular}[t]{@{}c@{}}Camera\\ users\end{tabular}} \\
 & \multicolumn{1}{c}{\%} & \multicolumn{1}{c}{\%} & \multicolumn{1}{c}{\%} & \multicolumn{1}{c}{\%} & \multicolumn{1}{c}{\%} & \multicolumn{1}{c}{\%} \\
\hline
\endfirsthead
\caption{Predictor variables by class of smartphone usage (\textit{continued}).} \\
\hline
Variables & \multicolumn{1}{c}{\begin{tabular}[t]{@{}c@{}}Advanced\\ users\end{tabular}} & \multicolumn{1}{c}{\begin{tabular}[t]{@{}c@{}}Broad non-\\ social-media \\ users\end{tabular}} & \multicolumn{1}{c}{\begin{tabular}[t]{@{}c@{}}Broad non-\\ commercial\\ users\end{tabular}} & \multicolumn{1}{c}{\begin{tabular}[t]{@{}c@{}}Basic\\ general\\ users\end{tabular}} & \multicolumn{1}{c}{\begin{tabular}[t]{@{}c@{}}Social media\\ and information\\ users\end{tabular}} & \multicolumn{1}{c}{\begin{tabular}[t]{@{}c@{}}Camera\\ users\end{tabular}} \\
 & \multicolumn{1}{c}{\%} & \multicolumn{1}{c}{\%} & \multicolumn{1}{c}{\%} & \multicolumn{1}{c}{\%} & \multicolumn{1}{c}{\%} & \multicolumn{1}{c}{\%} \\
\hline
\endhead
Class size & 39 & 7 & 24 & 12 & 10 & 8 \\
\hline
Frequency of   smartphone use &  &  &  &  &  &  \\
\hspace{3mm} Several times a day & 94 & 93 & 88 & 41 & 57 & 21 \\
\hspace{3mm} Every day & 5 & 7 & 11 & 35 & 27 & 18 \\
\hspace{3mm} Several times a week or less & 1 & 0 & 1 & 24 & 15 & 60 \\
Smartphone skills &  &  &  &  &  &  \\
\hspace{3mm} Advanced (5) & 58 & 36 & 31 & 10 & 9 & 8 \\
\hspace{3mm} Intermediate (4) & 33 & 46 & 50 & 38 & 20 & 12 \\
\hspace{3mm} Beginner (1-3) & 9 & 18 & 20 & 52 & 71 & 80 \\
Smartphone activities &  &  &  &  &  &  \\
\hspace{3mm} Browse websites & 100 & 98 & 100 & 83 & 82 & 31 \\
\hspace{3mm} Email & 99 & 97 & 94 & 68 & 65 & 40 \\
\hspace{3mm} Photo & 100 & 100 & 99 & 87 & 98 & 83 \\
\hspace{3mm} View social media & 100 & 0 & 100 & 16 & 100 & 6 \\
\hspace{3mm} Post social media & 92 & 0 & 71 & 3 & 77 & 2 \\
\hspace{3mm} Online purchase & 98 & 61 & 28 & 12 & 10 & 1 \\
\hspace{3mm} Online banking & 78 & 54 & 29 & 22 & 14 & 2 \\
\hspace{3mm} Install apps & 99 & 96 & 96 & 72 & 49 & 16 \\
\hspace{3mm} GPS & 98 & 97 & 92 & 72 & 53 & 24 \\
\hspace{3mm} Bluetooth & 72 & 58 & 43 & 32 & 8 & 6 \\
\hspace{3mm} Games & 85 & 61 & 61 & 39 & 38 & 24 \\
\hspace{3mm} Streaming & 98 & 78 & 70 & 36 & 34 & 0 \\
\hline
N & 1,019 & 179 & 641 & 321 & 261 & 200 \\
\hline
\end{longtable}
\end{center}

\newpage
\begin{center}
\begin{longtable}{lrrrrrrr}
\caption{\label{tab:table_D2} Sociodemographic and smartphone-related characteristics by class of smartphone usage.} \\
\hline
Variables & \multicolumn{1}{c}{\begin{tabular}[t]{@{}c@{}}Advanced\\ users\end{tabular}} & \multicolumn{1}{c}{\begin{tabular}[t]{@{}c@{}}Broad non-\\ social-media\\ users\end{tabular}} & \multicolumn{1}{c}{\begin{tabular}[t]{@{}c@{}}Broad non-\\ commercial\\ users\end{tabular}} & \multicolumn{1}{c}{\begin{tabular}[t]{@{}c@{}}Basic\\ general\\ users\end{tabular}} & \multicolumn{1}{c}{\begin{tabular}[t]{@{}c@{}}Social media\\ and information\\ users\end{tabular}} & \multicolumn{1}{c}{\begin{tabular}[t]{@{}c@{}}Camera\\ users\end{tabular}} & \multicolumn{1}{c}{p-Value} \\
 & \multicolumn{1}{c}{\%} & \multicolumn{1}{c}{\%} & \multicolumn{1}{c}{\%} & \multicolumn{1}{c}{\%} & \multicolumn{1}{c}{\%} & \multicolumn{1}{c}{\%} & \multicolumn{1}{l}{} \\
\hline
\endhead
Gender &  &  &  &  &  &  & \textless{}0.001 \\
\hspace{3mm} Female & 52 & 37 & 53 & 42 & 57 & 45 &  \\
\hspace{3mm} Male & 48 & 63 & 47 & 58 & 43 & 55 &  \\
Age &  &  &  &  &  &  & \textless{}0.001 \\
\hspace{3mm} 18-29 & 43 & 14 & 33 & 11 & 15 & 6 &  \\
\hspace{3mm} 30-39 & 28 & 22 & 21 & 19 & 21 & 12 &  \\
\hspace{3mm} 40-49 & 20 & 36 & 24 & 29 & 30 & 26 &  \\
\hspace{3mm} 50-59 & 7 & 22 & 18 & 27 & 25 & 41 &  \\
\hspace{3mm} 60+ & 2 & 6 & 4 & 15 & 8 & 15 &  \\
Educational attainment &  &  &  &  &  &  & \textless{}0.001 \\
\hspace{3mm} No high school degree & 42 & 36 & 44 & 54 & 62 & 70 &  \\
\hspace{3mm} High school degree & 34 & 28 & 31 & 18 & 21 & 15 &  \\
\hspace{3mm} College degree & 24 & 36 & 25 & 28 & 16 & 16 &  \\
Operating system &  &  &  &  &  &  & \textless{}0.001 \\
\hspace{3mm} iOS & 29 & 26 & 22 & 14 & 14 & 12 &  \\
\hspace{3mm} Android & 67 & 65 & 74 & 73 & 74 & 77 &  \\
\hspace{3mm} Other operating system & 5 & 8 & 4 & 13 & 12 & 11 & \\
\hline
\end{longtable}
\end{center}

\section{Sample 2}
\vspace{-8mm}
\setcounter{table}{0} \renewcommand{\thetable}{E.\arabic{table}}
\begin{center}
\begin{longtable}{lrrrrrr}
\caption{\label{tab:table_E1} Predictor variables by class of smartphone usage.} \\
\hline
Variables & \multicolumn{1}{c}{\begin{tabular}[t]{@{}c@{}}Advanced\\ users\end{tabular}} & \multicolumn{1}{c}{\begin{tabular}[t]{@{}c@{}}Broad non-\\ social-media \\ users\end{tabular}} & \multicolumn{1}{c}{\begin{tabular}[t]{@{}c@{}}Broad non-\\ commercial\\ users\end{tabular}} & \multicolumn{1}{c}{\begin{tabular}[t]{@{}c@{}}Basic\\ general\\ users\end{tabular}} & \multicolumn{1}{c}{\begin{tabular}[t]{@{}c@{}}Social media\\ and information\\ users\end{tabular}} & \multicolumn{1}{c}{\begin{tabular}[t]{@{}c@{}}Camera\\ users\end{tabular}} \\
 & \multicolumn{1}{c}{\%} & \multicolumn{1}{c}{\%} & \multicolumn{1}{c}{\%} & \multicolumn{1}{c}{\%} & \multicolumn{1}{c}{\%} & \multicolumn{1}{c}{\%} \\
\hline
\endfirsthead
\caption{Predictor variables by class of smartphone usage (\textit{continued}).} \\
\hline
Variables & \multicolumn{1}{c}{\begin{tabular}[t]{@{}c@{}}Advanced\\ users\end{tabular}} & \multicolumn{1}{c}{\begin{tabular}[t]{@{}c@{}}Broad non-\\ social-media \\ users\end{tabular}} & \multicolumn{1}{c}{\begin{tabular}[t]{@{}c@{}}Broad non-\\ commercial\\ users\end{tabular}} & \multicolumn{1}{c}{\begin{tabular}[t]{@{}c@{}}Basic\\ general\\ users\end{tabular}} & \multicolumn{1}{c}{\begin{tabular}[t]{@{}c@{}}Social media\\ and information\\ users\end{tabular}} & \multicolumn{1}{c}{\begin{tabular}[t]{@{}c@{}}Camera\\ users\end{tabular}} \\
 & \multicolumn{1}{c}{\%} & \multicolumn{1}{c}{\%} & \multicolumn{1}{c}{\%} & \multicolumn{1}{c}{\%} & \multicolumn{1}{c}{\%} & \multicolumn{1}{c}{\%} \\
\hline
\endhead
Class size & 17 & 8 & 19 & 25 & 12 & 20 \\
\hline
Frequency of smartphone use &  &  &  &  &  &  \\
\hspace{3mm} Several times a day & 98 & 96 & 94 & 69 & 60 & 18 \\
\hspace{3mm} Every day & 2 & 4 & 6 & 19 & 24 & 25 \\
\hspace{3mm} Several times a week or less & 0 & 0 & 0 & 13 & 15 & 57 \\
Smartphone skills &  &  &  &  &  &  \\
\hspace{3mm} Advanced (5) & 72 & 39 & 21 & 11 & 11 & 4 \\
\hspace{3mm} Intermediate (4) & 23 & 43 & 52 & 32 & 19 & 6 \\
\hspace{3mm} Beginner (1-3) & 5 & 19 & 27 & 56 & 70 & 90 \\
Smartphone activities &  &  &  &  &  &  \\
\hspace{3mm} Browse websites & 100 & 99 & 95 & 84 & 61 & 30 \\
\hspace{3mm} Email & 98 & 97 & 91 & 76 & 62 & 45 \\
\hspace{3mm} Photo & 100 & 96 & 99 & 92 & 84 & 72 \\
\hspace{3mm} View social media & 100 & 0 & 99 & 0 & 93 & 2 \\
\hspace{3mm} Post social media & 76 & 0 & 46 & 0 & 39 & 0 \\
\hspace{3mm} Online purchase & 93 & 75 & 37 & 14 & 9 & 1 \\
\hspace{3mm} Online banking & 70 & 54 & 25 & 15 & 6 & 2 \\
\hspace{3mm} Install apps & 97 & 80 & 62 & 30 & 4 & 1 \\
\hspace{3mm} GPS & 97 & 92 & 79 & 61 & 30 & 8 \\
\hspace{3mm} Bluetooth & 77 & 69 & 42 & 24 & 6 & 3 \\
\hspace{3mm} Games & 60 & 40 & 33 & 22 & 21 & 11 \\
\hspace{3mm} Streaming & 93 & 77 & 48 & 26 & 15 & 2 \\
\hline
N & 363 & 166 & 419 & 549 & 254 & 435 \\
\hline
\end{longtable}
\end{center}

\newpage
\begin{center}
\begin{longtable}{lrrrrrrr}
\caption{\label{tab:table_E2} Sociodemographic characteristics by class of smartphone usage.} \\
\hline
Variables & \multicolumn{1}{c}{\begin{tabular}[t]{@{}c@{}}Advanced\\ users\end{tabular}} & \multicolumn{1}{c}{\begin{tabular}[t]{@{}c@{}}Broad non-\\ social-media\\ users\end{tabular}} & \multicolumn{1}{c}{\begin{tabular}[t]{@{}c@{}}Broad non-\\ commercial\\ users\end{tabular}} & \multicolumn{1}{c}{\begin{tabular}[t]{@{}c@{}}Basic\\ general\\ users\end{tabular}} & \multicolumn{1}{c}{\begin{tabular}[t]{@{}c@{}}Social media\\ and information\\ users\end{tabular}} & \multicolumn{1}{c}{\begin{tabular}[t]{@{}c@{}}Camera\\ users\end{tabular}} & \multicolumn{1}{c}{p-Value} \\
 & \multicolumn{1}{c}{\%} & \multicolumn{1}{c}{\%} & \multicolumn{1}{c}{\%} & \multicolumn{1}{c}{\%} & \multicolumn{1}{c}{\%} & \multicolumn{1}{c}{\%} & \multicolumn{1}{l}{} \\
\hline
\endhead
Gender &  &  &  &  &  &  & \textless{}0.001 \\
\hspace{3mm} Female & 44 & 34 & 54 & 43 & 57 & 57 &  \\
\hspace{3mm} Male & 56 & 66 & 46 & 57 & 43 & 43 &  \\
Age &  &  &  &  &  &  &  \textless{}0.001 \\
\hspace{3mm} 18-27 & 29 & 7 & 17 & 4 & 8 & 1 &  \\
\hspace{3mm} 28-37 & 32 & 36 & 23 & 15 & 8 & 5 &  \\
\hspace{3mm} 38-47 & 21 & 19 & 21 & 18 & 17 & 10 &  \\
\hspace{3mm} 48-57 & 14 & 22 & 26 & 28 & 33 & 31 &  \\
\hspace{3mm} 58+ & 4 & 17 & 14 & 34 & 33 & 53 &  \\
Educational attainment &  &  &  &  &  &  & \textless{}0.001 \\
\hspace{3mm} No high school degree & 31 & 26 & 41 & 44 & 57 & 67 &  \\
\hspace{3mm} High school degree & 33 & 22 & 27 & 21 & 21 & 13 &  \\
\hspace{3mm} College degree & 36 & 52 & 32 & 34 & 22 & 21 & \\
\hline
\end{longtable}
\end{center}
\end{landscape}

\section{Sample 3}
\vspace{-8mm}
\setcounter{table}{0} \renewcommand{\thetable}{F.\arabic{table}}
\begin{center}
\begin{longtable}{lrrrr}
\caption{\label{tab:table_F1} Predictor variables by class of smartphone usage.} \\
\hline
Variables & \multicolumn{1}{c}{\begin{tabular}[t]{@{}c@{}}Advanced\\ users\end{tabular}} & \multicolumn{1}{c}{\begin{tabular}[t]{@{}c@{}}Broad non-\\ commercial\\ users\end{tabular}} & \multicolumn{1}{c}{\begin{tabular}[t]{@{}c@{}}Basic\\ general\\ users\end{tabular}} & \multicolumn{1}{c}{\begin{tabular}[t]{@{}c@{}}Camera\\ users\end{tabular}} \\
 & \multicolumn{1}{c}{\%} & \multicolumn{1}{c}{\%} & \multicolumn{1}{c}{\%} & \multicolumn{1}{c}{\%} \\
\hline
\endhead
Class size & 47 & 26 & 13 & 14 \\
\hline
Frequency of smartphone use &  &  &  &  \\
\hspace{3mm} Several times a day & 91 & 66 & 65 & 14 \\
\hspace{3mm} Every day & 7 & 23 & 28 & 13 \\
\hspace{3mm} Several times a week or less & 2 & 11 & 6 & 73 \\
Smartphone skills &  &  &  &  \\
\hspace{3mm} Advanced (5) & 64 & 26 & 26 & 7 \\
\hspace{3mm} Intermediate (4) & 28 & 38 & 44 & 23 \\
\hspace{3mm} Beginner (1-3) & 8 & 36 & 30 & 70 \\
Smartphone activities &  &  &  &  \\
\hspace{3mm} Browse websites & 100 & 96 & 100 & 47 \\
\hspace{3mm} Email & 99 & 86 & 89 & 48 \\
\hspace{3mm} Photo & 100 & 98 & 91 & 88 \\
\hspace{3mm} View social media & 100 & 100 & 4 & 19 \\
\hspace{3mm} Post social media & 89 & 69 & 8 & 4 \\
\hspace{3mm} Online purchase & 94 & 36 & 58 & 6 \\
\hspace{3mm} Online banking & 80 & 22 & 49 & 7 \\
\hspace{3mm} Install apps & 99 & 80 & 88 & 23 \\
\hspace{3mm} GPS & 98 & 73 & 81 & 32 \\
\hspace{3mm} Bluetooth & 76 & 37 & 54 & 8 \\
\hspace{3mm} Games & 87 & 57 & 58 & 16 \\
\hspace{3mm} Streaming & 97 & 66 & 60 & 12 \\
\hline
N & 566 & 315 & 159 & 173 \\
\hline
\end{longtable}
\end{center}

\begin{table}[H]
\caption{\label{tab:table_F2} Sociodemographic and smartphone-related characteristics by class of smartphone usage.}
\centering
\begin{tabular}{lrrrrr}
\hline
Variables & \multicolumn{1}{c}{\begin{tabular}[t]{@{}c@{}}Advanced\\ users\end{tabular}} & \multicolumn{1}{c}{\begin{tabular}[t]{@{}c@{}}Broad non-\\ commercial\\ users\end{tabular}} & \multicolumn{1}{c}{\begin{tabular}[t]{@{}c@{}}Basic\\ general\\ users\end{tabular}} & \multicolumn{1}{c}{\begin{tabular}[t]{@{}c@{}}Camera\\ users\end{tabular}} & \multicolumn{1}{c}{p-Value} \\
 & \multicolumn{1}{c}{\%} & \multicolumn{1}{c}{\%} & \multicolumn{1}{c}{\%} & \multicolumn{1}{c}{\%} & \multicolumn{1}{l}{} \\
\hline
Gender &  &  &  &  & 0.098 \\
\hspace{3mm} Female & 51 & 53 & 43 & 45 &  \\
\hspace{3mm} Male & 49 & 47 & 57 & 55 &  \\
Age &  &  &  &  & \textless{}0.001 \\
\hspace{3mm} 18-29 & 43 & 25 & 14 & 6 &  \\
\hspace{3mm} 30-39 & 29 & 20 & 21 & 9 &  \\
\hspace{3mm} 40-49 & 18 & 28 & 29 & 34 &  \\
\hspace{3mm} 50-59 & 9 & 21 & 25 & 32 &  \\
\hspace{3mm} 60+ & 1 & 5 & 11 & 20 &  \\
Educational attainment &  &  &  &  & \textless{}0.001 \\
\hspace{3mm} No high school degree & 43 & 56 & 47 & 55 &  \\
\hspace{3mm} High school degree & 35 & 23 & 23 & 19 &  \\
\hspace{3mm} College degree & 23 & 21 & 30 & 26 &  \\
Operating system &  &  &  &  & \textless{}0.001 \\
\hspace{3mm} iOS & 32 & 19 & 20 & 14 &  \\
\hspace{3mm} Android & 65 & 73 & 68 & 71 &  \\
\hspace{3mm} Other operating system & 3 & 8 & 11 & 15 & \\
\hline
\end{tabular}
\end{table}

\section{Sample 4}
\vspace{-8mm}
\setcounter{table}{0} \renewcommand{\thetable}{G.\arabic{table}}
\begin{table}[H]
\caption{\label{tab:table_G1} Predictor variables by class of smartphone usage.}
\centering
\begin{tabular}{lrrrr}
\hline
Variables & \multicolumn{1}{c}{\begin{tabular}[t]{@{}c@{}}Advanced\\ users\end{tabular}} & \multicolumn{1}{c}{\begin{tabular}[t]{@{}c@{}}Broad non-\\ social-media\\ users\end{tabular}} & \multicolumn{1}{c}{\begin{tabular}[t]{@{}c@{}}Social media\\ and information\\ users\end{tabular}} & \multicolumn{1}{c}{\begin{tabular}[t]{@{}c@{}}Camera\\ users\end{tabular}} \\
 & \multicolumn{1}{c}{\%} & \multicolumn{1}{c}{\%} & \multicolumn{1}{c}{\%} & \multicolumn{1}{c}{\%} \\
\hline
Class size & 39 & 19 & 19 & 23 \\
\hline
Frequency of smartphone use &  &  &  &  \\
\hspace{3mm} Several times a day & 92 & 78 & 63 & 27 \\
\hspace{3mm} Every day & 8 & 11 & 30 & 25 \\
\hspace{3mm} Several times a week or less & 0 & 11 & 7 & 48 \\
Smartphone skills &  &  &  &  \\
\hspace{3mm} Advanced (5) & 64 & 24 & 13 & 7 \\
\hspace{3mm} Intermediate (4) & 26 & 42 & 33 & 12 \\
\hspace{3mm} Beginner (1-3) & 10 & 34 & 54 & 82 \\
Smartphone activities &  &  &  &  \\
\hspace{3mm} Browse websites & 97 & 97 & 90 & 60 \\
\hspace{3mm} Email & 96 & 94 & 65 & 53 \\
\hspace{3mm} Photo & 100 & 93 & 92 & 91 \\
\hspace{3mm} View social media & 100 & 26 & 100 & 16 \\
\hspace{3mm} Post social media & 80 & 0 & 69 & 2 \\
\hspace{3mm} Online purchase & 72 & 55 & 19 & 2 \\
\hspace{3mm} Online banking & 70 & 47 & 30 & 8 \\
\hspace{3mm} Install apps & 98 & 76 & 38 & 29 \\
\hspace{3mm} GPS & 96 & 83 & 60 & 48 \\
\hspace{3mm} Bluetooth & 72 & 58 & 24 & 16 \\
\hspace{3mm} Games & 55 & 45 & 19 & 8 \\
\hspace{3mm} Streaming & 91 & 71 & 56 & 28 \\
\hline
N & 244 & 119 & 123 & 146 \\
\hline
\end{tabular}
\end{table}

\begin{table}[H]
\caption{\label{tab:table_G2} Sociodemographic characteristics by class of smartphone usage.}
\centering
\begin{tabular}{lrrrrr}
\hline
Variables & \multicolumn{1}{c}{\begin{tabular}[t]{@{}c@{}}Advanced\\ users\end{tabular}} & \multicolumn{1}{c}{\begin{tabular}[t]{@{}c@{}}Broad non-\\ social-media\\ users\end{tabular}} & \multicolumn{1}{c}{\begin{tabular}[t]{@{}c@{}}Social media\\ and information\\ users\end{tabular}} & \multicolumn{1}{c}{\begin{tabular}[t]{@{}c@{}}Camera\\ users\end{tabular}} & \multicolumn{1}{c}{p-Value} \\
 & \multicolumn{1}{c}{\%} & \multicolumn{1}{c}{\%} & \multicolumn{1}{c}{\%} & \multicolumn{1}{c}{\%} & \multicolumn{1}{l}{} \\
\hline
Gender &  &  &  &  & 0.132 \\
\hspace{3mm} Female & 46 & 50 & 58 & 55 &  \\
\hspace{3mm} Male & 54 & 50 & 42 & 45 &  \\
Age &  &  &  &  & \textless{}0.001 \\
\hspace{3mm} 18-29 & 49 & 25 & 28 & 3 &  \\
\hspace{3mm} 30-49 & 37 & 42 & 38 & 31 &  \\
\hspace{3mm} 50+ & 14 & 33 & 33 & 66 &  \\
Educational attainment &  &  &  &  & 0.001 \\
\hspace{3mm} No high school degree & 44 & 47 & 60 & 61 &  \\
\hspace{3mm} High school degree & 56 & 53 & 40 & 39 & \\
\hline
\end{tabular}
\end{table}

\end{document}